\documentclass[aps,graphicx,amsmath,amssymb,superscriptaddress,reprint,prl]{revtex4-2}

\usepackage{graphicx}
\usepackage{dcolumn}
\usepackage{bm}
\usepackage[colorlinks,linkcolor=blue,citecolor=blue]{hyperref}
\usepackage[normalem]{ulem}
\usepackage{xcolor}

\begin{document}

\title{Dipolar-driven mean-field criticality in the ferrimagnet Eu$_2$MnSi$_2$O$_7$}

\author{Masahiro Kawamata}
\affiliation{Department of Physics, Tohoku University, Sendai, Miyagi 980-8578, Japan}
\affiliation{Department of Physics, Tokyo Metropolitan University, Hachioji, Tokyo 156-0057, Japan}
\author{Maxim Avdeev}
\affiliation{Australian Centre for Neutron Scattering, Australian Nuclear Science and Technology Organisation, Kirrawee DC, NSW 2232, Australia}
\affiliation{School of Chemistry, The University of Sydney, Sydney, NSW 2006, Australia}
\author{Yusuke Nambu}
\email{nambu.yusuke.7s@kyoto-u.ac.jp}
\affiliation{Institute for Integrated Radiation and Nuclear Science, Kyoto University, Kumatori, Osaka 590-0494, Japan}
\affiliation{FOREST, Japan Science and Technology Agency, Kawaguchi, Saitama 332-0012, Japan}

\date{\today}

\begin{abstract}
We report mean-field critical behavior in Eu$_2$MnSi$_2$O$_7$, a melilite-type ferrimagnet with spin-only Eu$^{2+}$ and Mn$^{2+}$ moments and negligible orbital contributions.
Magnetization measurements combined with neutron powder diffraction reveal critical exponents close to the mean-field values, indicating that long-range dipolar interactions govern the asymptotic critical behavior in this insulating ferrimagnet.
The refined magnetic structure, described by the magnetic space group $P2_12_1^\prime2^\prime$, exhibits a tilted ferrimagnetic configuration driven by the Dzyaloshinskii-Moriya interaction, reflecting the noncentrosymmetric nature of the lattice.
These results extend the applicability of mean-field theory to complex insulating magnets and establish Eu$_2$MnSi$_2$O$_7$ as a platform for exploring ferrimagnetism and long-range interactions.
To our knowledge, this is the first insulating ferrimagnet in which dipolar interactions drive mean-field criticality.
\end{abstract}


\maketitle



Phase transitions are fundamental to a wide range of systems, from liquid-gas transitions in water to ferromagnetism~\cite{Aharoni2000}, superconductivity~\cite{Bardeen1957}, and the superfluidity of helium~\cite{Kapitza1938,Allen1938}.
Although their microscopic origins differ, these transitions share a key feature: collective changes in the macroscopic state of matter in response to external parameters such as temperature, pressure, or magnetic field.
A central concept for describing this behavior is universality, which groups seemingly distinct systems according to their common critical properties~\cite{Stanley1971}.

In magnetic systems, universality classes are determined primarily by the dimensionality of space and spin when short-range interactions dominate, typically expressed as $J(r)\sim e^{-r/b}$ with $b$ as the spatial scaling factor~\cite{Fischer2002}.
Under these conditions, ferromagnets, ferrimagnets, and antiferromagnets fall into the same universality class when their spatial and spin dimensionalities coincide~\cite{Jongh1974}.

When long-range interactions dominate, universality depends not only on dimensionality but also on the interaction range~\cite{Fisher1972}.
Accounting for such interactions is essential for describing critical behavior, especially in systems with itinerant electrons or large magnetic moments.
However, it remains unclear whether ferromagnets, ferrimagnets, and antiferromagnets belong to the same universality class as in the short-range case.
In metallic magnets, long-range correlations arise through the Ruderman-Kittel-Kasuya-Yosida (RKKY) interaction~\cite{Ruderman1954,Kasuya1956,Yosida1957}, an isotropic exchange mediated by conduction electrons that decays as $J(r)\sim r^{-(d+\sigma)}$, where $d$ is the dimensionality of the system and $\sigma$ characterizes the interaction range~\cite{Sak1973,Larson2010,Horita2017}.
As a result, many metallic systems exhibit critical exponents intermediate between those predicted by the Heisenberg model and the mean-field approximation~\cite{Yang2021}.

In contrast, the RKKY interaction is absent in insulators, where long-range coupling could originate instead from magnetic dipole-dipole (MDD) interactions.
In classical electromagnetism~\cite{Jackson1999}, an electric dipole moment is defined by two opposite charges; analogously, two hypothetical magnetic monopoles define a magnetic dipole moment.
Each magnetic moment produces a field that couples to its neighbors, leading to an interaction of the form
\begin{align}
\mathcal{H}_{\rm dip} = \frac{\mu_0}{4\pi|\mathbf r|^3}\sum_{i,j}\left[\mathbf m_i\cdot\mathbf m_j - 3(\mathbf m_i\cdot\hat{\mathbf r})(\mathbf m_j\cdot\hat{\mathbf r})\right],
\label{eq0}
\end{align}
which governs critical behavior driven by long-range dipolar interactions in insulating magnets~\cite{Aharony1973,Bruce1974,Ried1995}.
Within a mean-field framework, metals naturally exhibit long-range correlations via the RKKY interaction, whereas in insulators the generic long-range channel is the MDD coupling.
In what follows we therefore focus on mean-field expectations for systems with long-range interactions.

For ferromagnetic insulators with large total angular momentum $J=|L+S|$, such as GdCl$_3$ (Gd$^{3+}$; $J=S=7/2$)~\cite{Kotzler1973}, Dy(C$_2$H$_5$SO$_4$)$_3\cdot9$H$_2$O (Dy$^{3+}$; $J=15/2$)~\cite{Frowein1976}, LiTbF$_4$ (Tb$^{3+}$; $J=6$)~\cite{Beauvillain1980}, and LiHoF$_4$ (Ho$^{3+}$; $J=8$)~\cite{Bitko1996}, the observed critical exponents are close to mean-field values.
These results can be explained by incorporating dipolar interactions into the theoretical framework~\cite{Larkin1969}.
To date, dipolar-driven mean-field criticality in insulating magnets has been firmly established only for ferromagnets.
Establishing universality, however, requires coverage across ferro-, ferri- and antiferromagnets.
To our knowledge the ferrimagnetic and antiferromagnetic cases have remained unexplored.
Here we demonstrate that Eu$_2$MnSi$_2$O$_7$ provides the first ferrimagnetic example, thereby bridging this gap.

Here, we examine whether the same universality class applies to ferromagnets and ferrimagnets.
We focus on the melilite-type compound Eu$_2$MnSi$_2$O$_7$.
Its magnetization saturates at $2 \times (7/2 \times 2 - 5/2)
=9~\mu_{\rm B}$/f.u., corresponding to the difference between Eu$^{2+}$ ($S=7/2$) and Mn$^{2+}$ ($S=5/2$) moments~\cite{Endo2010,Toyoda2023}.
Previous studies suggest that Eu$_2$MnSi$_2$O$_7$ adopts a ferrimagnetic structure, with Eu and Mn spins oriented in opposite directions.

In melilite-type compounds with antiferromagnetic interactions, canting arises from the Dzyaloshinskii-Moriya (DM) interaction associated with noncentrosymmetry~\cite{Nambu2024,Kawamata2025}.
Eu$_2$MnSi$_2$O$_7$ also lacks inversion symmetry and is therefore expected to host a noncollinear magnetic structure.
However, its microscopic magnetic structure has not been resolved by neutron diffraction because Eu strongly absorbs neutrons ($\sigma_{\rm abs}=4530$~barn)~\cite{Sears1992}.

Here we report the synthesis of polycrystalline Eu$_2$MnSi$_2$O$_7$, combined with magnetization measurements and neutron powder diffraction analysis of its crystal and magnetic structures.
By determining the critical exponents through multiple methods, we show that this ferrimagnetic material exhibits behavior consistent with the mean-field model, as expected when MDD interactions dominate.

\begin{figure}[t!]
    \centering
    \includegraphics[width=0.85\linewidth]{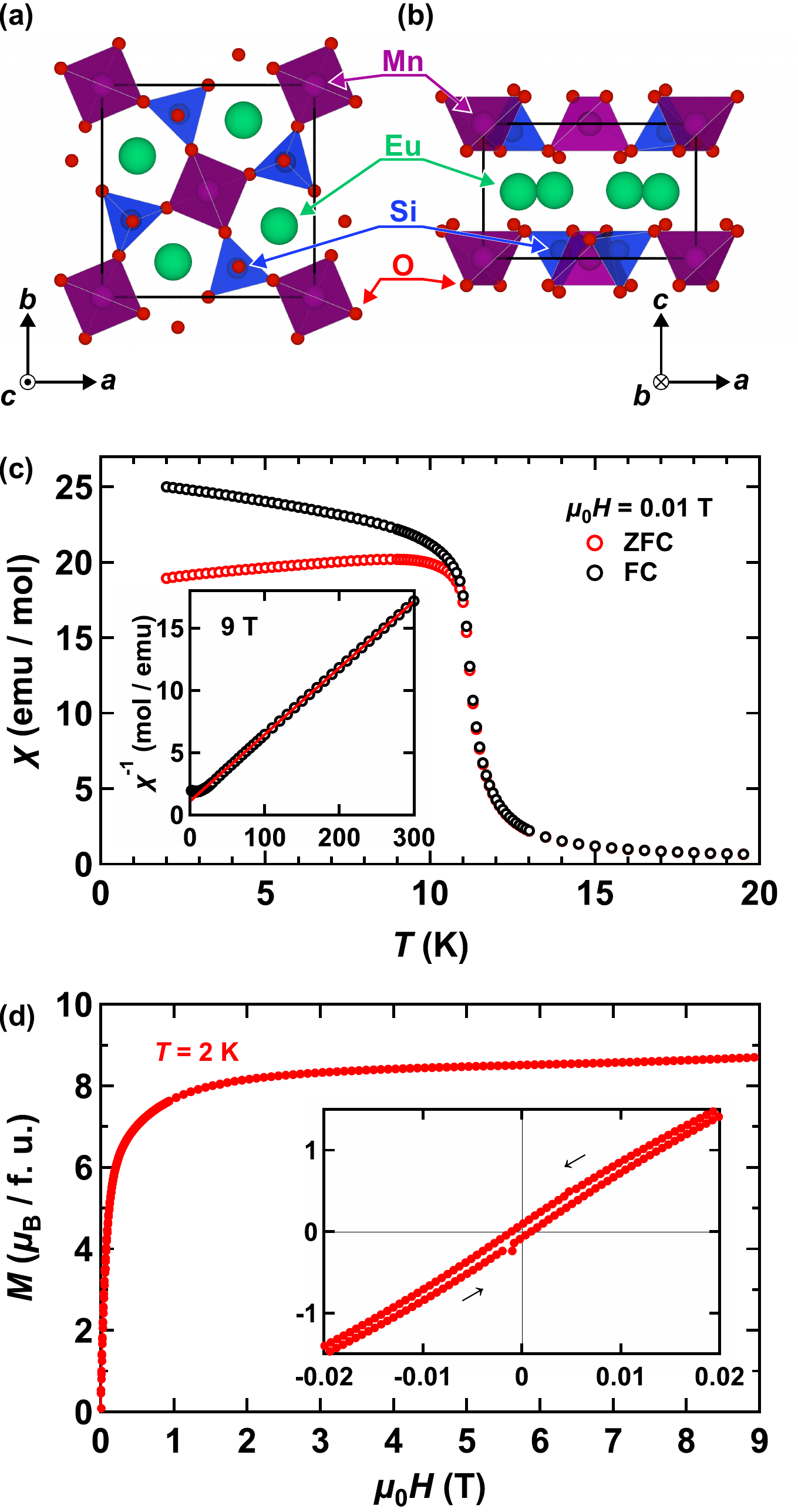}
    \caption{Crystal structure of the melilite compound Eu$_2$MnSi$_2$O$_7$ with the space group $P\bar{4}2_1m$ projected onto (a) the $ab$- and (b) the $ac$-plane. (c) Temperature dependence of the magnetic susceptibility at $\mu_0H=0.01$~T. Inset: inverse susceptibility under 9~T with a Curie-Weiss fit (red line). (d) Field dependence of the isothermal magnetization. Inset: low-field region at 2~K showing increasing and decreasing field sweeps.}
\label{f1}
\end{figure}


Polycrystalline samples of Eu$_2$MnSi$_2$O$_7$ were prepared by a solid-state reaction.
Magnetization and neutron powder diffraction (NPD) measurements were used to determine the magnetic properties, magnetic structure, and critical exponents; experimental details are given in the End Matter.



Figure~\ref{f1}(c) shows the temperature dependence of the magnetic susceptibility $\chi \equiv M/H$ for Eu$_2$MnSi$_2$O$_7$ under an applied field of $\mu_0H = 0.01$~T.
A ferromagnetic-like transition is evident from the sharp increase around 12~K in both zero-field-cooled (ZFC) and field-cooled (FC) protocols, followed by a clear bifurcation between them.
The inverse susceptibility $\chi^{-1}$ under 9~T [inset to Fig.~\ref{f1}(c)] increases linearly above 100~K.
To minimize the effects of residual magnetic fields and domain formation at low fields, as well as to reduce uncertainties in the high-temperature regime, the Curie-Weiss fitting was performed using high-field data.
Fitting the data with the Curie-Weiss law, $\chi = C/(T-\theta_{\rm W})$, in the range 100--300~K yields an effective magnetic moment of $\mu_{\rm eff} = 12.218(4)~\mu_{\rm B}$ and a Weiss temperature of $\theta_{\rm W} = -20.4(1)$~K.

The field dependence of the isothermal magnetization is presented in Fig.~\ref{f1}(d).
At the lowest temperature (2~K), the magnetization reaches 8.6~$\mu_{\rm B}$ per formula unit at 9~T.
This behavior is consistent with the previous report~\cite{Endo2010}, which proposed a ferrimagnetic structure with Eu and Mn moments oriented in opposite directions.
Such an assignment is reasonable, given that the dominant interaction in Eu$_2$MnSi$_2$O$_7$ is antiferromagnetic, as reflected by $\theta_{\rm W}<0$.
In addition, a very small coercivity is observed as $H \rightarrow 0$ [inset to Fig.~\ref{f1}(d)].
The detailed magnetic structure will be refined from neutron diffraction data, as discussed below.


\begin{figure*}[t!]
    \centering
    \includegraphics[width=0.85\linewidth]{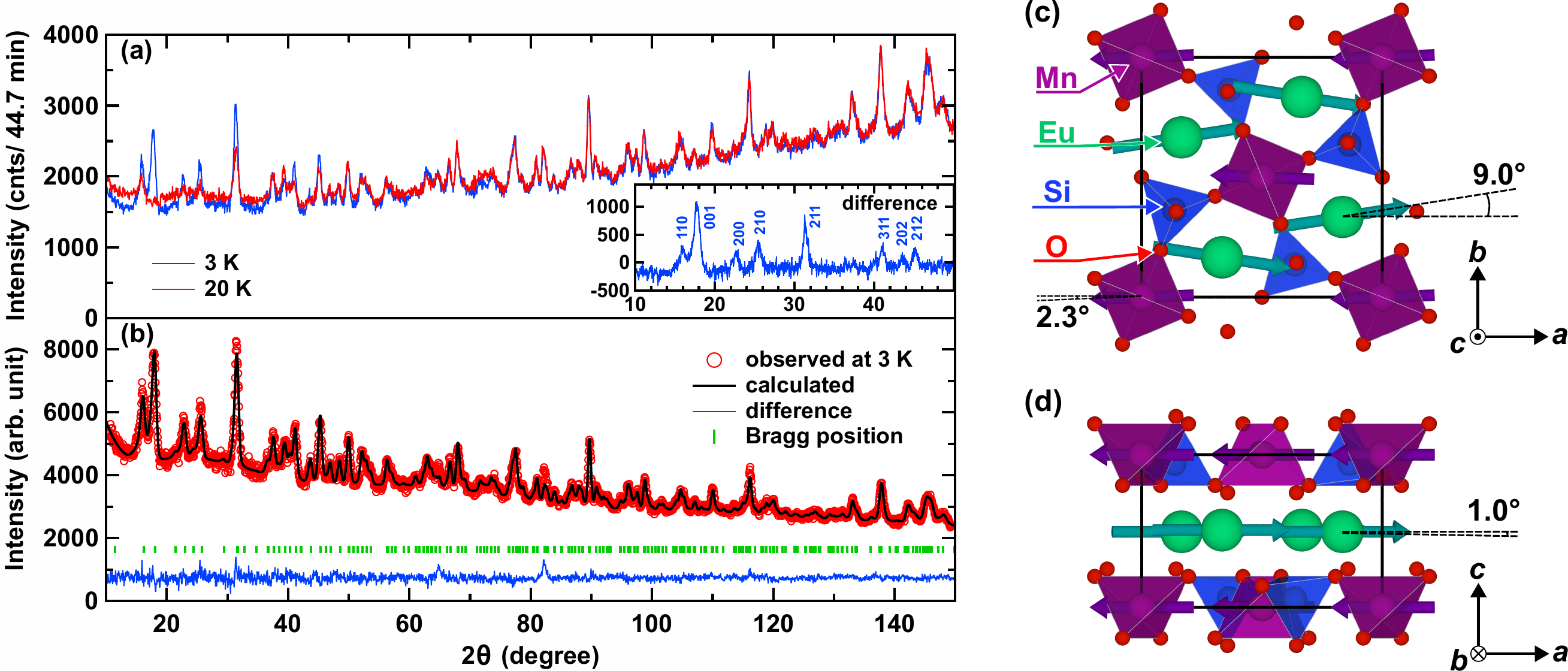}
    \caption{(a) Neutron powder diffraction data at 3 and 20~K ($\lambda=1.622$~Å). Inset: difference profile between the two temperatures. (b) Rietveld refinement of the 3~K pattern ($\chi^2=1.36$). Red circles: observed; black line: calculated; blue line: difference; green ticks: Bragg positions of Eu$_2$MnSi$_2$O$_7$. (c,d) Refined magnetic structure with space group $P2_12_1^{\prime}2^{\prime}$ projected onto (c) the $ab$ plane and (d) the $ac$ plane.}
\label{f2}
\end{figure*}

Figure~\ref{f2}(a) presents raw neutron powder diffraction data from Echidna collected at 3 and 20~K, corresponding to below and above the magnetic transition temperature.
The differences between the two diffraction profiles are attributed to magnetic reflections with wavevector $\vec{q}_{\rm m}=(0,0,0)$~r.l.u. [inset to Fig.~\ref{f2}(a)].
The nonmagnetic phase at 20~K, together with the crystal structure refinement, is shown in Fig.~S1~\cite{suppl}, where neutron absorption corrections were applied.
The structure is well described by the tetragonal space group $P\bar{4}2_1m$, and refined parameters are listed in Table~S1~\cite{suppl}.
As schematically illustrated in Figs.~\ref{f1}(a,b), Mn and Si atoms are each surrounded by four oxygen atoms, and corner-sharing MnO$_4$ and SiO$_4$ tetrahedra form a planar network within the $ab$-plane.
Mn atoms occupy a two-dimensional square lattice, which stacks along the $c$ axis separated by Eu atoms.

To refine the magnetic structure, we analyzed the data using magnetic space groups (MSGs), as detailed in the End Matter. 
Testing all candidates revealed that $P2_12_1^{\prime}2^{\prime}$, with a reliability factor of $R_{\rm mag}=6.45$\%, best reproduces the experimental data [Fig.~\ref{f2}(c,d)].
A previous second-harmonic generation study reported the magnetic point group $2^{\prime}2^{\prime}2$~\cite{Toyoda2023}, in good agreement with our findings.
The second-best fit was obtained for $Cm^{\prime}m2^{\prime}$ ($R_{\rm mag}=7.99$\%), while the other models failed to reproduce the data.
A combined Rietveld analysis of nuclear and magnetic reflections was carried out; details of the symmetry lowering and structural parameters are given in the End Matter and Table~S1~\cite{suppl}.

The only observed thermodynamic anomaly appears in the temperature dependence of the magnetic susceptibility.
Below the transition, the moments of both Eu$^{2+}$ and Mn$^{2+}$ are required to refine the NPD data, indicating that they order simultaneously at $T_{\rm C}=10.6$~K.
At 3~K, the ordered moments are 7.3(2) and 5.0(2)~$\mu_{\rm B}$ for Eu$^{2+}$ and Mn$^{2+}$, respectively, close to the expected values of 7~$\mu_{\rm B}$ for Eu$^{2+}$ ($4f^7$) and 5~$\mu_{\rm B}$ for Mn$^{2+}$ ($3d^5$).
Representative MDD energy scales and their relation to the exchange scale are discussed in the End Matter.

The refined moment components are $m_x^{\rm Eu}=7.23(14)$, $m_y^{\rm Eu}=1.14(17)$, $m_z^{\rm Eu}=0.12(3.15)$~$\mu_{\rm B}$, and $m_x^{\rm Mn}=-4.98(16)$, $m_y^{\rm Mn}=-0.20(1.41)$, $m_z^{\rm Mn}=0$~$\mu_{\rm B}$.
While the moments primarily lie in the $ab$-plane, the deviations, particularly $m_z^{\rm Eu}$ and $m_y^{\rm Mn}$, carry large statistical uncertainties and cannot be determined precisely from powder data.
A full refinement of the tilt angle and space group symmetry will require single-crystal diffraction under applied fields, which lies beyond the present scope.

Nevertheless, approximate tilt angles can be inferred.
The Mn$^{2+}$ moment lies in the $ab$-plane with a ferromagnetic configuration, tilted $\pm 2.3^\circ$ from the $a$-axis [Fig.~\ref{f2}(c)].
The Eu$^{2+}$ moment shows a larger in-plane tilt of $\pm 9.0^\circ$ [Fig.~\ref{f2}(c)], with a small out-of-plane component corresponding to a $\pm 1.0^\circ$ deviation [Fig.~\ref{f2}(d)].
Such tilts are typical in melilite-type compounds and originate from the DM interaction associated with the noncentrosymmetric structure, as also reported in Sr$_2$MnSi$_2$O$_7$~\cite{Nambu2024}, Ba$_2$CoGe$_2$O$_7$~\cite{Hutanu2014}, and Sr$_2$CoSi$_2$O$_7$~\cite{Dutta2023}.
A minimal spin-Hamiltonian estimate of the canting ratios is given in the End Matter.


We now determine the critical exponents from macroscopic magnetization and compare them with neutron results.
All relevant magnetic susceptibility and magnetization data are presented in Fig.~S3~\cite{suppl}.
For compensated antiferromagnets, critical exponents are commonly determined from the temperature dependence of magnetic reflections, often with $\vec{q}_{\rm m} \neq 0$, rather than from bulk magnetization~\cite{Hutanu2014,Dutta2023,Sazonov2023}.
In Eu$_2$MnSi$_2$O$_7$, however, the antiferromagnetically coupled Eu and Mn moments do not cancel, yielding a finite ferrimagnetic component.
Owing to this finite component, the critical exponents can be determined directly from macroscopic magnetization measurements.
The Kouvel-Fisher, modified-Arrott, and scaling analyses used below are summarized in the End Matter.

\begin{figure*}[t!]
    \centering
    \includegraphics[width=0.85\linewidth]{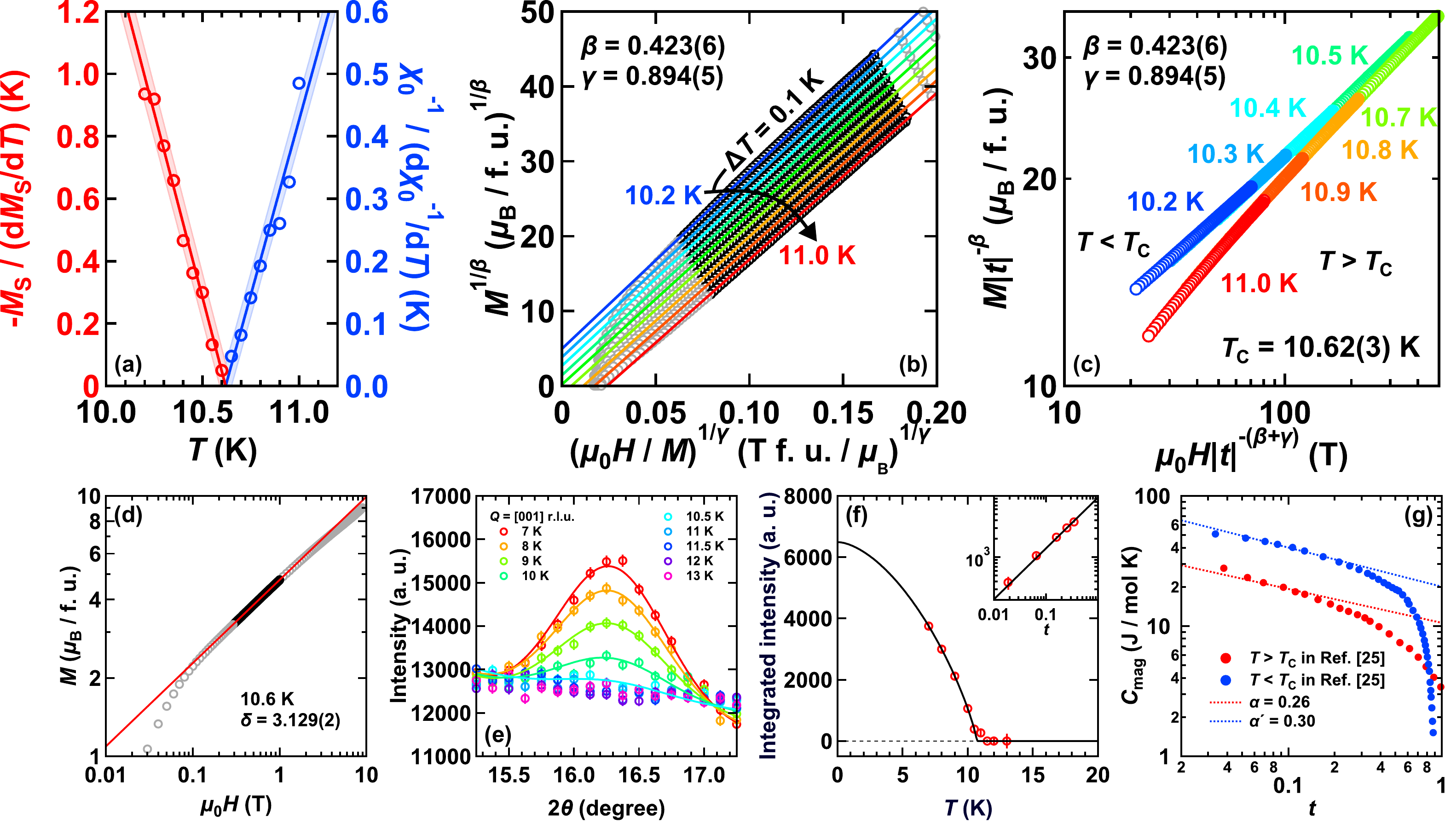}
    \caption{(a) Kouvel-Fisher plot for spontaneous magnetization $M_S(T)$ and inverse initial susceptibility $\chi_0^{-1}$. Linear fits are shown as red and blue lines; shaded areas denote uncertainties in $\beta$, $\gamma$, and $T_{\rm C}$. (b) Modified Arrott plot for $10.2\leq T\leq11.0$~K and $0.3\leq\mu_0H\leq1$~T. Black circles: fitted data; grey circles: data outside the fitting range. (c) Scaled magnetization as a function of reduced field above and below $T_{\rm C}$. (d) Critical isotherm at 10.6~K (dots) with fit (red line) yielding $\delta=3.129(2)$. (e) Temperature dependence of the magnetic reflection at $\vec{Q}=(0,0,1)$. (f) Integrated intensity at $\vec{Q}=(0,0,1)$ with fit giving $\beta=0.41(4)$. (g) Magnetic specific heat $C_{\rm mag}$ (solid circles) compared with fits using $\alpha=0.26$ and $\alpha=0.31$ (red and blue lines) from Ref.~\cite{Endo2010}.
}
\label{f3}
\end{figure*}


Critical exponents were determined by linear fits of the Kouvel-Fisher plots near $T_{\rm C}$ [Fig.~\ref{f3}(a)], followed by construction of the generalized modified Arrott plot [Fig.~\ref{f3}(b)].
The transition temperature was determined to be $T_{\rm C}=10.62(3)$~K, mostly consistent with the 10.7~K reported previously~\cite{Endo2010}.
The resulting exponents are $\beta=0.423(6)$ and $\gamma=0.894(5)$ for $10.2\leq T \leq 11.0$~K and $0.3\leq \mu_0H \leq 1$~T.


The reliability of these values can be confirmed by scaling theory.
Defining reduced variables $m=M|t|^{-\beta}$ and $h=\mu_0H|t|^{-(\beta+\gamma)}$, all curves should collapse onto two universal branches above and below $T_{\rm C}$.
As shown in Fig.~\ref{f3}(c), the data scale well using $\beta=0.423(6)$, $\gamma=0.894(5)$, and $T_{\rm C}=10.62(3)$~K, confirming their consistency.


The exponent $\delta$ was determined by two methods.
First, the Widom relation $\delta = 1 + \gamma/\beta$~\cite{Widom1965} gives $\delta=3.11(3)$.
Second, a log-log analysis of $M$ versus $\mu_0H$ at $T_{\rm C}$ yields $\delta=3.129(2)$ [Fig.~\ref{f3}(d)], in excellent agreement with the Widom estimate.


The temperature dependence of the magnetic reflection from 7 to 13~K is depicted in Fig.~\ref{f3}(e).
The integrated intensity [Fig.~\ref{f3}(f)] follows $I \propto |t|^{2\beta}$, yielding $T_{\rm C}=10.7(1)$~K and $\beta=0.41(4)$.
These values are consistent with those obtained from macroscopic measurements.

We tested the validity of the above estimates of $\alpha$ against the zero-field specific-heat data reported previously~\cite{Endo2010}.
Applying the Rushbrooke and Griffiths relations gives $\alpha=0.26(1)$ and $\alpha^{\prime}$ in the range 0.25(2)--0.30(13), as detailed in the End Matter.
Figure~\ref{f3}(g) shows that fits using $\alpha=0.26$ and $\alpha^{\prime}=0.30$ reproduce the data near $T_{\rm C}$.

\begin{figure}[t!]
    \centering
    \includegraphics[width=0.85\linewidth]{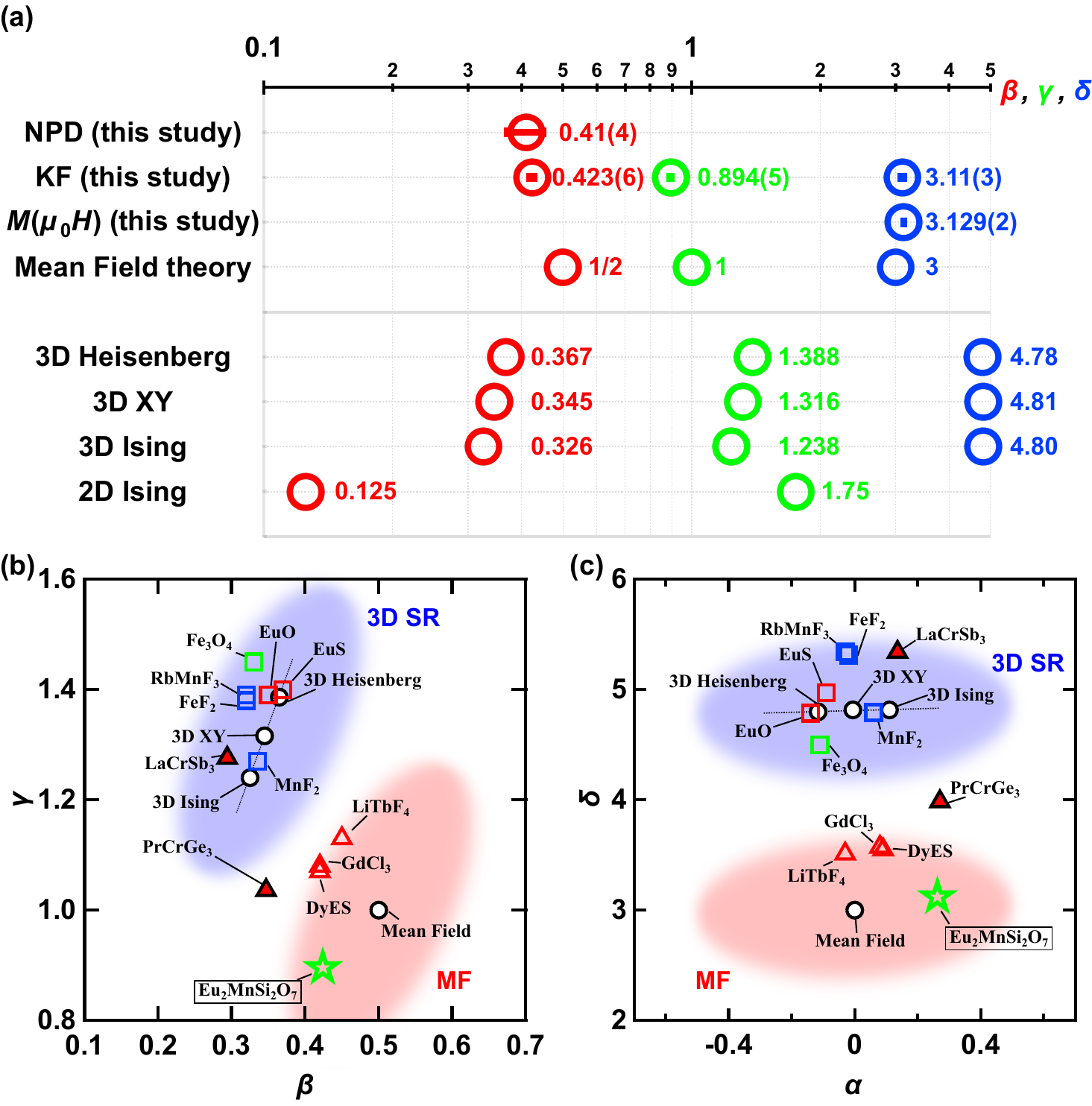}
    \caption{(a) Critical exponents $\beta$, $\gamma$, and $\delta$ of Eu$_2$MnSi$_2$O$_7$ compared with model values~\cite{Privman1991}. (b,c) $\beta$ vs $\gamma$ and $\alpha$ vs $\delta$. Open circles: theoretical values for mean-field, 3D-Heisenberg, 3D-XY, and 3D-Ising models~\cite{Privman1991}. Star: Eu$_2$MnSi$_2$O$_7$ (this work). Squares: ferromagnets (red)~\cite{Menyuk1971,Nielsen1971}, ferrimagnets (purple)~\cite{Olav1982}, and antiferromagnets (blue)~\cite{Sabba1981,Heller1966,Mattsson1994} with SR interactions. Triangles: ferromagnets with RKKY (filled) and MDD (open) interactions~\cite{Beauvillain1980,Kotzler1973,Frowein1976,Yang2021,Yang2022}. Shaded regions denote 3D-SR (red) and MF (blue) universality classes.}
\label{f4}
\end{figure}

Figures~\ref{f4}(a-c) summarize the critical exponents of Eu$_2$MnSi$_2$O$_7$ alongside representative models and materials.
When short-range interactions dominate, the exponents for ferromagnets (EuO~\cite{Menyuk1971,Nielsen1971}, EuS~\cite{Nielsen1971}), ferrimagnets (Fe$_3$O$_4$~\cite{Olav1982}), and antiferromagnets (RbMnF$_3$~\cite{Sabba1981}, MnF$_2$~\cite{Heller1966}, FeF$_2$~\cite{Mattsson1994}) depend on the spatial and spin dimensionalities.
By contrast, ferromagnets governed by long-range MDD interactions, such as LiTbF$_4$~\cite{Beauvillain1980}, GdCl$_3$~\cite{Kotzler1973}, and Dy(C$_2$H$_5$SO$_4$)$_3\cdot9$H$_2$O (DyES)~\cite{Frowein1976}, fall into the mean-field universality class.

Eu$_2$MnSi$_2$O$_7$, though ferrimagnetic, also approximates the mean-field universality class.
As an insulating compound with a charge gap of 1.23~eV at the $\Gamma$ point~\cite{Toyoda2023}, Eu$_2$MnSi$_2$O$_7$ lacks RKKY interactions; instead, MDD coupling provides the relevant long-range interaction channel owing to the large Eu and Mn moments.
Further discussion of interaction-range considerations and antiferromagnetic analogues is given in the End Matter.

In summary, we have established mean-field-like critical behavior in the insulating ferrimagnet Eu$_2$MnSi$_2$O$_7$.
Magnetization, scaling analyses, and neutron powder diffraction consistently yield critical exponents close to the mean-field values.
Neutron diffraction further reveals a tilted ferrimagnetic structure described by the magnetic space group $P2_12_1^\prime2^\prime$, with Eu and Mn moments ordering simultaneously.
Although exchange interactions stabilize the ferrimagnetic ordered state, the asymptotic critical behavior is governed by long-range magnetic dipole-dipole interactions.
Eu$_2$MnSi$_2$O$_7$ therefore provides a ferrimagnetic platform for studying dipolar-controlled criticality in insulating magnets.

\begin{acknowledgments}
We thank T. Arima, H. Kawamura, Y. Onose, and Y. Tabata for their valuable discussions.
This work was supported by JSPS KAKENHI (Grant Nos. 21H03732, 22H05145, 24K00572, 25K01489, 22J22265, 22KJ0312, 26H02027), the JST FOREST (Grant No. JPMJFR202V), and the Graduate Program in Spintronics at Tohoku University.
\end{acknowledgments}

\clearpage
\section*{End Matter}

\setcounter{equation}{0}
\renewcommand{\theequation}{A\arabic{equation}}

\textit{Experimental details.---}
Eu$_2$O$_3$, MnO, Si, and SiO$_2$ were mixed in a stoichiometric ratio, sealed in a quartz ampoule, then reacted at 1100$^{\circ}$C for 48~hours.
X-ray powder diffraction was used to evaluate the sample quality, and no secondary phases were detected.
Magnetization data were collected using the Physical Property Measurement System (Quantum Design) in the temperature range of 2 to 300~K with applied magnetic fields up to $\mu_0 H=9$~T.
The samples were packed into capsules, and analysis was performed taking into account the demagnetization factor corresponding to their shape.
The critical exponents were obtained by the Kouvel-Fisher (KF) plot~\cite{Kouvel1964} and scaling analysis using isothermal magnetization data.
NPD experiments were performed using the high-resolution diffractometer Echidna~\cite{Avdeev2018} and high-intensity diffractometer Wombat~\cite{Studer2006} at ANSTO, Australia.
The incident neutron wavelengths were $\lambda=1.622$ and 1.54~{\AA} for Echidna and Wombat, respectively.
To minimize effects from the high neutron absorber Eu, the powder sample was packed in a double-wall cylindrical cell.

\textit{Magnetic-structure analysis.---}
Six maximal subgroups are allowed by symmetry.
Among these, $P\bar{4}^{\prime}2_1 m^{\prime}$ (\#113.270) and $P\bar{4}^{\prime}2_1^{\prime}m$ (\#113.269) forbid magnetic moments on Mn sites due to the $\bar{4}^\prime..$ site symmetry at the $2a$ position, and were therefore excluded.
The four remaining candidates were $P\bar{4}2_1^{\prime}m^{\prime}$ (\#113.271), $P\bar{4}2_1 m.1$ (\#113.267), $Cm^{\prime}m2^{\prime}$ (\#35.167), and $P2_12_1^{\prime}2^{\prime}$ (\#18.19).
Their atomic arrangements and moment configurations are summarized in Figs.~S2(a-e)~\cite{suppl}.
For refinement of the 3~K profile, the space group $P2_12_12$ was adopted, and the corresponding structural parameters are listed in Table~S1~\cite{suppl}.
The nuclear structure was described in $P2_12_12$, and the magnetic ordering was treated within the MSG $P2_12_1^{\prime}2^{\prime}$.
The O3 site splits into two crystallographically independent positions, while the displacements of other atoms are minimal.
These distortions may be underestimated due to spherical averaging inherent in powder data.
To clarify symmetry lowering associated with the magnetic transition, future studies using single-crystal diffraction with field-aligned domains will be required.

Using the observed magnetic moments and interionic separations in Eq.~\ref{eq0}, we estimate representative pairwise MDD energies of 0.37, 0.62, and 0.082~K for Eu-Mn, Eu-Eu, and Mn-Mn, respectively.
By comparison, $|\theta_{\rm W}|$ = 20.4(1)~K provides an experimental measure of the collective exchange scale and is substantially larger than the MDD energies.
This comparison concerns characteristic energy scales and does not constitute a microscopic determination of individual exchange constants.
Exchange interactions thus stabilize the ferrimagnetic ordered state, whereas the MDD interaction can govern the asymptotic critical behavior through its long-range character.

\textit{Estimate of the canting.---}
The small canting can be quantified using a minimal spin Hamiltonian including exchange and DM terms,
\begin{align}
\mathcal{H} = \sum_{\langle i,j\rangle}\left\{J_{ij}{\mathbf S}_i\cdot {\mathbf S}_j + {\mathbf D}_{ij} \cdot \left({\mathbf S}_i\times {\mathbf S}_j\right)\right\},
\end{align}
where $\langle i,j\rangle$ denotes interacting spin pairs.
For simplicity, we retain only nearest-neighbor Eu-Mn interactions and approximate the Eu moments as confined to the $ab$-plane.
The Mn-Eu1/Eu2 and Mn-Eu3/Eu4 bonds are parameterized by $(J, D_z)$ and $(J^{\prime}, D^{\prime}_z)$, respectively.
For the first set of bonds, the interaction energy is
\begin{align}
E_1 = 4J {\mathbf S}_{\rm Mn} \cdot {\mathbf S}_{\rm Eu1} + 4D_z ({\mathbf S}_{\rm Mn} \times {\mathbf S}_{\rm Eu1})_z,
\end{align}
with an analogous expression for the second set.
Minimizing these energies with respect to the spin angles gives
\begin{align}
\frac{D_z}{J}&=\tan(\theta_{\rm Eu}+\theta_{\rm Mn}),\\
\frac{D_z^{\prime}}{J^{\prime}}&=-\tan(\theta_{\rm Eu}-\theta_{\rm Mn}).
\end{align}
With $\theta_{\rm Eu}=9.0^{\circ}$ and $\theta_{\rm Mn}=2.3^{\circ}$, we obtain $D_z/J\sim 0.20$ and $D_z^{\prime}/J^{\prime}\sim -0.12$.
These ratios provide again only order-of-magnitude estimates because spherical averaging of the powder data and possible domain formation limit the accuracy of the refined canting angles.

\textit{Critical-exponent analysis.---}
Using the reduced temperature $t \equiv (T-T_{\rm C})/T_{\rm C}$, the magnetization $M(t,H)$ near the critical point follows~\cite{Privman1991}
\begin{align}
M(t,H=0) &\propto |t|^\beta & (T<T_{\rm C}),\\
M(t=0,H) &\propto |\mu_0H|^{1/\delta} & (T=T_{\rm C}),\\
\chi(t,H=0)&\propto |t|^{-\gamma^\prime} & (T<T_{\rm C}),\\
\chi(t,H=0)&\propto t^{-\gamma} & (T>T_{\rm C}).
\end{align}
The Arrott-Noakes equation, valid in the asymptotic critical region, is expressed as~\cite{Arrott1967}
\begin{align}
\left(\frac{H}{M}\right)^{1/\gamma} = \frac{T-T_{\rm C}}{T_1} + \left(\frac{M}{M_1}\right)^{1/\beta},
\end{align}
where $T_1$ and $M_1$ are material-specific constants.
From the modified Arrott plots, $M_S(T)$ is obtained from the vertical intercepts below $T_{\rm C}$, and $\chi_0^{-1}(T)$ from the horizontal intercepts above $T_{\rm C}$.
Their derivatives satisfy the Kouvel-Fisher relations~\cite{Kouvel1964},
\begin{align}
M_S(T)\left(\frac{{\rm d}M_S(T)}{{\rm d}T}\right)^{-1} &= \frac{T-T_{\rm C}}{\beta(T)},\\
\chi^{-1}(T)\left(\frac{{\rm d}\chi^{-1}(T)}{{\rm d}T}\right)^{-1} &= \frac{T-T_{\rm C}}{\gamma(T)}.
\end{align}
The iterative procedure used 50 cycles, requiring convergence of $\beta$ and $\gamma$ to within 0.005, and the thermal stability of each isothermal curve was within 0.5~K, with a standard deviation of $\sigma\sim0.00085$~K.
Scaling theory predicts~\cite{Privman1991}
\begin{align}
M(\mu_0H,t) = |t|^\beta f_{\pm}\left(\mu_0H|t|^{-(\beta+\gamma)}\right),
\end{align}
where $f_{\pm}$ corresponds to $T \gtrless T_{\rm C}$.
The Widom relation $\delta = 1 + \gamma/\beta$~\cite{Widom1965} was used to check the critical isotherm, and the integrated intensity of the magnetic reflection was fitted by $I \propto |t|^{2\beta}$.
For the specific-heat comparison, the critical behavior was written as~\cite{Privman1991}
\begin{align}
C &\propto A^{\prime}|t|^{-\alpha^{\prime}} \quad (T<T_{\rm C}),\\
C &\propto A|t|^{-\alpha} \quad (T>T_{\rm C}),
\end{align}
with $A$ and $A^{\prime}$ being material-dependent constants.
Applying the values of $\beta$ and $\gamma$ from Fig.~\ref{f3}(c) to the Rushbrooke relation, $\alpha+2\beta+\gamma=2$~\cite{Rushbrooke1963}, yields $\alpha=0.26(1)$.
In addition, using $\beta$ and $\delta$ from Fig.~\ref{f3}(c,d,f) in Griffiths' equation, $\beta(1+\delta)=2-\alpha$~\cite{Griffiths1965}, gives $\alpha^{\prime}$ in the range 0.25(2)-0.30(13).

\textit{Universality comparison and outlook.---}
In general, the validity of the mean-field approximation depends on the dimensionality, the magnitude of the magnetic moments, and the coordination number.
However, when long-range interactions dominate the critical behavior, the relevant criterion is whether a spin can effectively couple to a sufficiently large number of neighbors within the interaction range.
In such cases, as long as the interactions extend over a sufficient range to establish three-dimensional correlations, mean-field-like criticality can emerge even in systems with moderate structural coordination numbers.

In metals, the dominant long-range interaction is RKKY.
The corresponding exchange $J(r)=r^{-(d+\sigma)}$ spans behavior from mean-field to short-range universality classes depending on $\sigma$~\cite{Fisher1972}.
The exponents approach mean-field as $\sigma\rightarrow d/2$ and approach short-range as $\sigma\rightarrow 2$.
In PrCrGe$_3$ ($d=3$, $\sigma\sim1.6$)~\cite{Yang2021}, the exponents lie between mean-field and short-range, whereas in LaCrSb$_3$ ($d=3$, $\sigma\sim1.9$)~\cite{Yang2022} they are closer to the 3D short-range values.

Eu$_2$MnSi$_2$O$_7$, though ferrimagnetic, also approximates the mean-field universality class.
We note that three-dimensional dipolar magnets can exhibit multiplicative logarithmic corrections to mean-field scaling close to $T_{\rm C}$~\cite{Aharony1973,Larkin1969}, but within our experimental resolution no systematic deviations from pure power-law behavior were detected.
This is expected in the limit of strong long-range interactions.
Microscopically, a ferrimagnet carries both a net magnetization and a staggered component arising from inequivalent sublattices.
As a result, the spin-wave spectrum generically splits into multiple branches, rather than a single ferromagnetic mode~\cite{Nambu2020}, while time-reversal symmetry is still broken globally.
Eu$_2$MnSi$_2$O$_7$ therefore has a finite uniform ferrimagnetic component together with internal antiferromagnetic correlations between sublattices.
As an insulating compound with a charge gap of 1.23~eV at the $\Gamma$ point~\cite{Toyoda2023}, Eu$_2$MnSi$_2$O$_7$ lacks RKKY interactions; instead, MDD coupling provides the relevant long-range interaction channel owing to the large Eu and Mn moments.

In dipolar systems, the MDD interaction remains finite in the long-wavelength limit ($Q\rightarrow0$), in contrast to short-range exchange interactions whose contribution vanishes~\cite{Aharony1973,Bruce1974}.
As a result, although the magnetic structure itself is primarily stabilized by exchange interactions, the critical behavior near the transition is governed by the long-range nature of the MDD interaction.
This separation of roles between exchange (short-range) and dipolar (long-range) interactions is well established in classical dipolar magnets, where dipolar interactions dominate the asymptotic criticality despite being weaker in microscopic energy scale.
In this context, the mean-field-like critical exponents observed in Eu$_2$MnSi$_2$O$_7$ naturally indicate that the critical behavior is controlled by dipolar interactions, even though the underlying ferrimagnetic order is stabilized by exchange coupling.

This work provides guidance for identifying mean-field criticality in the remaining magnetic class, antiferromagnets.
By introducing multiple magnetic ions in melilites, one can promote strong 3D correlations.
Eu$_2$MnSi$_2$O$_7$ already hosts a large $J=7/2$ moment on the $A$ site (Eu$^{2+}$).
We suggest two routes to realize antiferromagnetic analogues.
First, dilute Eu$^{2+}$ with non-magnetic Ca$^{2+}$, Sr$^{2+}$, or Ba$^{2+}$ to reduce the $A$-site moment towards $S\rightarrow 5/2$.
Alternatively, replace Eu$^{2+}$ with a magnetic ion of $S=5/2$.
Fabricating such systems would test whether antiferromagnets dominated by long-range interactions also fall into the mean-field universality class.
A positive outcome would extend the universality picture across ferromagnets, ferrimagnets, and antiferromagnets, sharpening our understanding of phase-transition universality.

\end{document}